\documentstyle[12pt]{article}
\begin{document}
\newcommand{\beq}{\begin{equation}}
\newcommand{\eeq}{\end{equation}}
\newcommand{\bea}{\begin{eqnarray}}
\newcommand{\eea}{\end{eqnarray}}
\newcommand{\half}{\frac{1}{2}}
\newcommand{\ehalf}{\frac{e}{2}}
\newcommand{\ihalf}{\frac{i}{2}}
\newcommand{\khalf}{\frac{\kappa}{2}}
\newcommand{\iehalf}{\frac{ie}{2}}
\newcommand{\ghalf}{\frac{g}{2}}
\newcommand{\quart}{\frac{1}{4}}
\newcommand{\kquart}{\frac{\kappa}{4}}
\newcommand{\G}{\Gamma}
\renewcommand{\O}{\Omega}
\renewcommand{\S}{\Sigma}
\newcommand{\U}{\Upsilon}
\renewcommand{\a}{\alpha}
\renewcommand{\b}{\beta}
\newcommand{\e}{\epsilon}
\newcommand{\g}{\gamma}
\newcommand{\k}{\kappa}
\renewcommand{\l}{\lambda}
\newcommand{\s}{\sigma}
\newcommand{\bs}{\bar\s}
\newcommand{\bss}{\bar\s\s}
\renewcommand{\t}{\theta}
\newcommand{\pdp}{\Phi^{\dag}\Phi}
\newcommand{\opar}{\bar\eta}
\newcommand{\ze}{\bar{z}}
\newcommand{\cA}{\cal A}
\newcommand{\cB}{\cal B}
\newcommand{\cH}{\cal H}
\newcommand{\dmu}{\partial_{\mu}}
\newcommand{\dmup}{\partial^{\mu}}
\newcommand{\aslash}{\not\!\!\cA}
\newcommand{\hslash}{\not\!\!\cH}
\newcommand{\Dslash}{\not\!\! D}
\newcommand{\dslash}{\not\!\partial}
\newcommand{\pau}{\vec{\tau}}
\newcommand{\tsq}{\theta^2}
\newcommand{\otsq}{\bar\theta^2}
\newcommand{\ts}{\bar\theta\theta}
\newcommand{\tu}{\t\sigma^{\mu}\tte}
\newcommand{\tte}{\bar\theta}
\newcommand{\qeqe}{\{\bar{{\cal Q}}[\eta_{\pm}],{\cal 
                   Q}[\eta_{\pm}]\}\vert}
\def\hepth#1{ {\tt hep-th/#1}}
\begin{titlepage}
\begin{flushright}
{ ~}\vskip -1in
US-FT/12-98\\
\hepth{9807002}\\
\end{flushright}
\vspace*{31pt}

\def\today{\ifcase\month\or
January\or February\or March\or April\or May\or June\or
July\or August\or September\or October\or November\or December\fi,
\number\year}
\centerline{\Large \bf Self-dual solitons in N=2 supersymmetric}
\vskip 3mm
\centerline{\Large \bf semilocal Chern-Simons theory}
\vskip 13mm
\centerline{\sc Jos\'e D. Edelstein}  
\vskip 5mm
\centerline{{\it Depto. de F\'{\i}sica de Part\'{\i}culas,
Universidade de Santiago de Compostela}}
\centerline{{\it E-15706 Santiago de Compostela, Spain}}
\vskip 2mm
\centerline{{\tt edels@fpaxp1.usc.es}}
\vskip 17mm
\centerline{\sc Abstract}
\vskip 9mm

\noindent
We embed the semilocal Chern-Simons-Higgs theory into an $N=2$
supersymmetric system. We construct the corresponding conserved
supercharges and derive the Bogomol'nyi equations of the model
from supersymmetry considerations. We show that these equations
hold provided certain conditions on the coupling constants as well
as on the Higgs potential of the system, which are a consequence of 
the huge symmetry of the theory, are satisfied. They admit 
string-like solutions which break one half of the supersymmetries
--BPS Chern-Simons semilocal cosmic strings-- whose magnetic flux
is concentrated at the center of the vortex. We study such 
solutions and show that their stability is provided by supersymmetry 
through the existence of a lower bound for the energy, even though 
the manifold of the Higgs vacuum does not contain non-contractible 
loops. 
\vfill

\end{titlepage}
\newpage
\setcounter{footnote}{0}

In recent years semilocal vortex solutions \cite{VA} have attracted 
a wide attention in connection with both cosmological problems 
\cite{JP} as well as the physics of planar condensed matter 
systems \cite{K}. 
In this last context, the possible relevance of soliton solutions 
was particularly stressed in the case of Chern-Simons theories
where charged vortices are known to be, in fact, charged anyons
\cite{FM}.
The main feature of the so-called semilocal vortex configurations 
is that they are stable --with finite energy per unit length-- for
certain values of the parameters, even when the vacuum manifold is 
simply connected \cite{RAL}. 
They arise in models with a Lagrangian density having both global 
and local symmetries where the global symmetry is longer than the
local one which also has to be completely broken. 
These kinds of models were shown to admit stable string solutions even 
though the manifold of minima for the potential energy does not 
contain non-contractible loops \cite{VA}. 
The stability is provided by the requirement that the gradient energy 
density falls off sufficiently fast at infinity.

~

Vortices emerging as finite energy solutions of gauge theories usually
satisfy a topological bound for the energy, the so-called Bogomol'nyi 
bound \cite{Bogo}. 
It has been observed that Bogomol'nyi bounds seem to reflect the 
presence of an underlying extended supersymmetric structure 
\cite{WO}-\cite{ENS}.
In particular, for gauge theories with spontaneous symmetry breaking 
and a topological charge, admitting of an $N=1$ supersymmetric version, 
it was shown that the N=2 supersymmetric extension, which requires 
certain conditions on coupling constants as well as on the Higgs
potential, has a central charge coinciding with the topological charge 
\cite{HS,HS2,ENS}.
The existence of a Bogomol'nyi bound in such theories is a corollary
of this statement.

~

In spite of the fact that semilocal vortices appear in topologically
trivial models, it was recently shown that the Bogomol'nyi bound 
satisfied by the semilocal cosmic string, first obtained in 
Ref.\cite{VA}, can be also traced up to come from supersymmetry
\cite{EN}.
In the gravitationally coupled system, semilocal cosmic string and 
multi-string solutions were explicitly found and studied a few years 
ago \cite{GORRS}. 
The semilocal cosmic string can be consistently embedded into an 
$N=2$ supergravity theory whenever a critical relation between 
coupling constants takes place and its stability is a consequence 
of the underlying $N=2$ supercharge algebra \cite{JDE}. 
Although the geometry produced by the semilocal cosmic string is
asymptotically conical, Killing spinors can be consistently defined
provided the field configuration saturates the Bogomol'nyi bound.
Nevertheless, as the would-be fermionic Nambu-Goldstone zero modes 
generated by the action of broken supercharges are non-normalizable, 
there is no remnant supersymmetry in the physical 
Hilbert space \cite{JDE}. 
This gives a non-trivial example of Witten's claim on the vanishing 
of the cosmological constant without Bose-Fermi degeneracy in $2+1$ 
dimensions \cite{W1}. It is interesting to point here that, in 
certain $2+1$ systems, it is also possible to have solitons that
saturate the Bogomol'nyi bound even when the cosmological constant
does not vanish \cite{KK}.

~

In this letter we shall study the supersymmetric 
embedding of semilocal vortices that appear in Chern-Simons
gauge theory. 
We will consider the semilocal Chern-Simons-Higgs (CSH) system,
which is given by the following Lagrangian density
\beq
{\cal L}_{CSH} = \kquart \e^{\mu\nu\l}F_{\mu\nu}A_{\l} + 
(D_{\mu}\Phi)^{\dagger}(D^{\mu}\Phi) - V_{CSH}(\pdp) ~,
\label{lagcsh}
\eeq
where $\Phi$ is a complex Higgs doublet and the covariant derivative 
is $D_{\mu} = \dmu + ieA_{\mu}$. The metric is choosen to be $g^{\mu\nu}
= (+ - -)$, and the specific form of the symmetry breaking potential 
$V_{CSH}$ will be determined below by supersymmetry arguments.
The classical dynamics of the system is governed by the corresponding
equations of motion
\beq
\kappa {\cal H}^{\mu} = e{\cal J}^{\mu} ~,
\label{motion1}
\eeq
\beq
D_{\mu}D^{\mu}\Phi = - 
\frac{\delta{V}_{CSH}(\pdp)}{\delta\Phi^{\dag}} ~,
\label{motion2}
\eeq
where ${\cal H}^{\mu}$ is the dual field strength ${\cal H}^{\mu} =
\half\e^{\mu\nu\l}F_{\nu\l}$ and ${\cal J}^{\mu}$ is the Higgs
current
\beq
-i{\cal J}^{\mu} = \Phi^{\dag}(D^{\mu}\Phi) - 
(D^{\mu}\Phi)^{\dag}\Phi ~.
\label{current}
\eeq
The Lagrangian density (\ref{lagcsh}) has a global $SU(2)$ symmetry as 
well as a local $U(1)$ invariance, under which the Higgs field changes 
as $\Phi \to e^{i\a(x)}\Phi$.
When the Higgs field acquires a definite vacuum expectation value
the symmetry is broken to a global $U(1)$. 
The vacuum manifold is the three-sphere $|\Phi| = v$, which has no 
non-contractible loops. 
However, as the gradient energy density must fall off sufficiently
fast asymptotically, fields at infinity ought to lie on a gauge
orbit, that is, a circle lying on the three-sphere.

~

Let us start by considering the construction of the
supersymmetric semilocal CSH model 
whose bosonic sector is described by the Lagrangian density presented 
in Eq.(\ref{lagcsh}). The minimal $N=1$ supersymmetric extension of 
this model is given by an action which in superspace reads:
\beq
{\cal L}_{N=1}^{CSH} = \int d^2\theta \left[ \khalf 
\bar\O\G + \half (\bar{\cal D} + ie\bar\G)\U^{\dag}({\cal D} 
- ie\G)\U + {\cal W}(\U^{\dag}\U) \right] ~,
\label{lagsusy2}
\eeq
where $\U$ is a complex doublet superfield $\U \equiv (\Phi,\Psi,F)$
and $\G$ is the spinor gauge superfield which in the Wess-Zumino gauge 
reads $\G \equiv (A_{\mu},\rho)$. $\O$ is the corresponding superfield 
strength, ${\cal W}$ is the superpotential that defines the system
and ${\cal D}$ is the usual supercovariant derivative:
\beq
{\cal D} = \partial_{\tte} + i\tte\g^{\mu}\dmu ~. 
\label{super}
\eeq
It must be stressed that $\rho$ is a Majorana fermion while the 
Higgsino doublet $\Psi$ is a Dirac spinor. $F$ is an auxiliary 
complex doublet. Finally, the $\gamma$-matrices are taken to be in the 
Majorana representation $\gamma^0 = \tau^3$, $\gamma^1 = i\tau^1$ 
and $\gamma^2 = -i\tau^2$, that is, $\g^{\mu}\g^{\nu} = g^{\mu\nu}
+ i\e^{\mu\nu\l}\g_{\l}$.

~

Written in component fields, the preceding lagrangian takes the form
\bea
{\cal L}_{N=1}^{CSH} & = & \kquart \e^{\mu\nu\l}F_{\mu\nu}A_{\l} + 
(D_{\mu}\Phi)^{\dag}(D^{\mu}\Phi) + {\cal W}^{\prime}(\pdp)(F^{\dag}\Phi 
+ \Phi^{\dag}F) \nonumber \\
& + & F^{\dag}F + i\bar\Psi\Dslash\Psi - \khalf\bar\rho\rho 
- \half {\cal W}^{\prime\prime}(\pdp)(\Phi^{t}\Phi\bar\Psi\Psi^c
+ \Phi^{\dag}\Phi^*\bar\Psi^c\Psi) \nonumber \\
& + & ie(\bar\Psi\rho\Phi - \Phi^{\dag}\bar\rho\Psi) 
- [{\cal W}^{\prime}(\pdp) + \pdp{\cal 
W}^{\prime\prime}(\pdp)]\bar\Psi\Psi ~,
\label{susycsh}
\eea
where the supraindex $c$ in $\Psi^c$ stands for charge conjugation 
(complex conjugation in the Majorana representation) of $\Psi$.
Note that there is no D-term contribution to the potential energy
of the bosonic sector. This feature comes from the fact that we were 
not forced to introduce an additional real scalar superfield in the
system, that is, it is not necessary for our construction to have a 
Fayet-Iliopoulos term in the lagrangian. This is a consequence of the 
fact that the photino $\rho$ is not a dynamical field as can be
immediately seen in the above expression.
The Higgs potential of the theory then comes 
from an F-term contribution whose explicit form is obtained using the
algebraic equation of motion of the auxiliary field and results
to be
\beq
V_{CSH}(\pdp) = {\cal W}^{\prime}(\pdp)^2\pdp ~.
\label{ftermcsh}
\eeq
Since the photino $\rho$ is an auxiliary field in the supersymmetric
CSH system, we eliminate it by using its equation of motion
\beq
\rho = \frac{ie}{\kappa}\left({\Psi^t}^c\Phi - 
\Phi^{\dag}\Psi\right) ~,
\label{eqrho}
\eeq
where the transposition in ${\Psi^t}^c$ refers to its transformation
properties under $SU(2)$. The Lagragian of the system is then written
as
\bea
{\cal L}_{N=1}^{CSH} & = & \kquart \e^{\mu\nu\l}F_{\mu\nu}A_{\l} + 
(D_{\mu}\Phi)^{\dag}(D^{\mu}\Phi) - {\cal W}^{\prime}(\pdp)^2\pdp 
\nonumber \\ & + & i\bar\Psi\Dslash\Psi - \half \left({\cal 
W}^{\prime\prime}(\pdp) + \frac{e^2}{\kappa}\right)
(\Phi^{t}\Phi\bar\Psi\Psi^c + \Phi^{\dag}\Phi^*\bar\Psi^c\Psi) \nonumber \\
& - & \left[{\cal W}^{\prime}(\pdp) + \pdp\left({\cal 
W}^{\prime\prime}(\pdp) - 
\frac{e^2}{\kappa}\right)\right]\bar\Psi\Psi ~. 
\label{susycsh2}
\eea
The corresponding action is invariant under the following set 
of $N=1$ supersymmetry transformations with parameter $\eta$:
\beq 
\delta_{\eta} A_{\mu} = - 
\frac{e}{\k}\left(\bar\Psi^t\Phi\g_{\mu}\eta 
+ \opar\g_{\mu}\Phi^{\dag}\Psi\right) \;\;\; , \;\;\; 
\delta_{\eta} \Phi = \opar\Psi 
\label{trafos1}
\eeq
\beq 
\delta_{\eta} \Psi = \left[-i\gamma^{\mu}D_{\mu}\Phi - {\cal 
W}^{\prime}(\pdp)\Phi\right]\eta ~.
\label{trafos2}
\eeq

~

\noindent
Now, in order to impose the $N=2$ supersymmetric invariance of the 
theory, we should consider transformations with a complex parameter 
$\eta_c$ (an infinitesimal Dirac spinor), since this implies the 
existence of two supersymmetries \cite{LLW}. The above transformations 
(\ref{trafos1}) and (\ref{trafos2}) with a complex parameter $\eta_c = 
\eta e^{-i\alpha}$ are equivalent to the transformations with a real
parameter $\eta$ followed by a phase rotation for the fermions, $\Psi \to
e^{i\alpha}\Psi$.
Then, $N=2$ supersymmetry requires invariance under this fermion
rotation. It is immediate to see that the invariance of
(\ref{susycsh2}) under this transformation is achieved if and only if
\beq
{\cal W}^{\prime\prime}(\pdp) = - \frac{e^2}{\k} ~.
\label{conds2}
\eeq
That is, the model is invariant under an extended supersymmetry 
provided the condition (\ref{conds2}) is fulfilled. 
This kind of condition appears in general when, starting from an 
$N=1$ supersymmetric gauge model, one attempts to impose a second 
supersymmetry so as to accommodate different $N=1$ multiplets into 
an $N=2$ multiplet.

~

It is worthwhile to point out that this condition leads to a quartic
superpotential,
\beq
{\cal W}(\pdp) = - \frac{e^2}{2\k}\left(\pdp - v^2\right)^2 ~,
\label{superp}
\eeq
with $v^2$ a real constant, exactly as it happens in the local $N=2$ 
supersymmetric Abelian CSH system originally studied in Ref.\cite{LLW},
in spite of the fact that here we are dealing with a topologically 
trivial system in the sense discussed above.
We can then obtain the Higgs potential which, after (\ref{ftermcsh}), 
reads
\beq
V_{CSH}(\pdp) = \frac{e^4}{\k^2}\pdp\left(\pdp - 
v^2\right)^2 ~.
\label{higgscsh}
\eeq
The specific shape of the Higgs potential, as announced, has been 
dictated just by supersymmetry considerations. We have obtained, as 
a consequence of our requirement of extended supersymmetry, a sixth-order 
Higgs potential which has two degenerate ground states: the symmetric 
vacuum $\Phi = 0$ and a sphere of asymmetric states $\pdp = v^2$.
The same behaviour for the Higgs potential is also known to appear in 
the purely local systems, either in the Abelian case \cite{LLW} or in
the non-Abelian case \cite{K}. However, it is interesting to point out
that it is not a universal feature of Chern-Simons theories. Indeed,
it has been recently shown that certain self-dual solitons emerge in
N=2 supersymmetric (local and Abelian) Chern-Simons gauge theory for 
eight-order Higgs potentials \cite{GFMG}. We will come back to this
point below.

~

Let us write down explicitly the lagrangian density of the $N=2$ 
supersymetric semilocal CSH system:
\bea
{\cal L}_{N=2}^{CSH} & = & \kquart \e^{\mu\nu\l}F_{\mu\nu}A_{\l} + 
(D_{\mu}\Phi)^{\dag}(D^{\mu}\Phi) - 
\frac{e^4}{\k^2}\pdp\left(\pdp - v^2\right)^2 \nonumber \\
& + & i\bar\Psi\Dslash\Psi + \frac{e^2}{\k}\left(3\pdp - 
v^2\right)\bar\Psi\Psi ~. 
\label{susycsh3}
\eea
Our interest will be focused on purely bosonic field 
configurations of this system. 
We then introduce the following useful notation: Given a functional 
$\Xi$ depending both on bosonic and fermionic fields, we will use 
$\Xi\vert$ to refer to that functional evaluated in the purely bosonic 
background,
\beq
\Xi\vert \equiv \Xi\vert_{\S,\Psi=0} ~.
\label{spb}
\eeq
Under condition (\ref{spb}), the only non-vanishing $N=2$
supersymmetric transformation corresponding to this lagrangian
is that of the Higgsino $\Psi$,
\beq
\delta_{\eta_c}\Psi\vert = \left[-i\gamma^{\mu}D_{\mu}\Phi +
\frac{e^2}{\k}\left(\pdp - v^2\right)\Phi\right]\eta_c ~. 
\label{tratri}
\eeq

~

In order to study the reasons why the condition (\ref{conds2}),
that ensure $N=2$ supersymmetry, is also needed for the existence
of a Bogomol'nyi bound, let us analyse further the supercharge 
algebra. The conserved charge can be obtained following the Noether
method to be:
\beq
{\cal Q}[\eta_c] = \int d^2x \Psi^{\dag}\left[\g^{\mu}D_{\mu}\Phi 
+ i\frac{e^2}{\k}\left(\pdp - v^2\right)\Phi\right]\eta_c ~. 
\label{supcharge}
\eeq
Since we are interested in connecting the $N=2$ supercharge algebra 
with the Bogomol'nyi bound, we assume static configurations 
and we restrict ourselves to a purely bosonic solution of the theory 
after computing the algebra. Let us also impose, for reasons that will
be clear below, a `chirality' condition on the fermionic parameter
$\eta_c$, $\gamma^0\eta_{\pm} = \pm\eta_{\pm}$.

~

We now compute the supercharge algebra and obtain
\beq
\qeqe = (M \pm Z) \eta_{\pm}^{\dag}\eta_{\pm} ~,
\label{qeqe}
\eeq
with $M$ the mass of the purely bosonic configuration and $Z$ the 
central extension of the algebra whose explicit form is given by
\beq
Z = - \int d^2x \left[ e{\cal F}\pdp - \frac{e^2}{\kappa}{\cal J}_0
(\pdp - v^2) \right] ~,
\label{central}
\eeq
where we have used ${\cal F}$ for the magnetic field ${\cal F} = 
{\cal H}^0 = \e^{z\ze}F_{z\ze}$, and $\e^{z\ze}$ is the covariant 
antisymmetric tensor. After Gauss' law --the $0$-component 
of eq.(\ref{motion1})-- the mass of the configuration reads
\beq
M = \int d^2x \left[ \frac{\k^2}{4e^2}\frac{{\cal F}^2}{\pdp} 
+ (D_k\Phi)^{\dag}(D_k\Phi) + \frac{e^4}{\k^2}\pdp\left(\pdp 
- v^2\right)^2 \right] ~,
\label{mass1}
\eeq
whereas the central charge $Z$ results to be proportional to the 
magnetic flux $\Phi_{\mbox{m}}$ across the planar system:
\beq
Z = ev^2\Phi_{\mbox{m}} ~.
\label{topol1}
\eeq
This is one of the main points of our work: once the relation 
between the central charge and the topological charge (quantized
magnetic flux) is established, the Bogomol'nyi bound of the system,
\beq
M \geq ev^2|\Phi_{\mbox{m}}| ~,
\label{bog}
\eeq
comes from the fact that the supersymmetry algebra is positive 
definite. Indeed, the bracket given 
in (\ref{qeqe}) can be written as the integral of a fermionic bilinear, 
\beq
\{\bar{\cal Q}[\eta_{\pm}],{\cal Q}[\eta_{\pm}]\}| = \int d^2x 
(\delta_{\eta_{\pm}}\Psi)^{\dag}(\delta_{\eta_{\pm}}\Psi) \geq 0 ~,
\label{cota}
\eeq
in such a way that the lower bound is saturated whenever one half of the
supersymmetries is preserved by the field configuration,
$\delta_{\eta_{\pm}}\Psi = 0$. These are nothing but the Bogomol'nyi
equations of the model. Their explicit form can be obtained from 
(\ref{tratri}), after Gauss' law, resulting
\beq
D_z\Phi = 0 ~,
\label{bogo1}
\eeq
\beq
{\cal F} = \frac{2e^3}{\kappa^2}\pdp(\pdp - v^2) ~,
\label{bogo2}
\eeq
for the upper sign, whereas for the lower sign we obtain
\beq
D_{\ze}\Phi = 0 ~,
\label{abogo1}
\eeq
\beq
{\cal F} = - \frac{2e^3}{\kappa^2}\pdp(\pdp - v^2) ~.
\label{abogo2}
\eeq
These equations were originally found in Ref.\cite{K} for the bosonic 
system in a non-supersymmetric context. Let us remark that we have 
obtained them just by asking our configuration to have an unbroken 
supersymmetry.  Also, having originated from the supercharge algebra, 
the energy of the configurations that saturate the Bogomol'nyi
bound is expected to be quantum mechanically exact. 

~

Let us consider the case when the unbroken supersymmetry is that
corresponding to the chiral parameter $\eta_+$. This corresponds
to a bosonic field configuration that solves the set of equations
(\ref{bogo1}-\ref{bogo2}). Note that the expression (\ref{bogo2})
for the magnetic field in terms of the Higgs doublet is completely 
analogous to that of the selfdual abelian CSH
vortex solution \cite{HKP,JW}. In that case, it was deduced from the 
relation between ${\cal F}$ and the Higgs field that the magnetic field 
is concentrated in a cylindrical shell around the (n-th order) zero of 
the Higgs field. 
It is interesting to point out that the above remark is not true for 
the BPS semilocal CSH solution. To see this, it is enough to consider
the following ansatz (not the most general one, see Ref.\cite{MH}) for 
the Higgs doublet and the gauge field which mantains the axial symmetry 
and has non-trivial topology
\beq
\Phi(\s,\bs) = v\varphi(\bss)\bs\,\Phi_1 + v\omega(\bss)\,\Phi_2 ~,
\label{doublet}
\eeq
\beq
A_\s(\s,\bs) = -i\frac{\xi(\bss)}{e\s} ~,
\label{gaugef}
\eeq
where $(\s,\bs) = ev(z,\ze)$ are dimensionless coordinates, 
$\varphi(\bss)$, $\omega(\bss)$ and $\xi(\bss)$ are real functions 
and $\Phi_1$ and $\Phi_2$ give an orthonormal basis for the 
fundamental representation of the global $SU(2)$. The 
orthogonality of $\Phi_1$ and $\Phi_2$ ensures that the effect
of a spatial rotation can be removed from both components by a 
suitable $SU(2)\times{U(1)}$ symmetry transformation. It is
immediate to see that this solution has $n$ quanta of unit
magnetic flux
\beq
\Phi_{\mbox{m}} = 2\oint A_\s d\s = \frac{4\pi{n}}{e} ~,
\label{flux}
\eeq
provided $\xi(\bss) \to n$ when the radius $r=(\bss)^{1/2}$ goes to 
infinity. $n$ is an integer which is inmediately identified as the 
topological charge of the configuration.
The Higgs doublet should asymptotically approach a minimum 
of the potential. For that reason, $\varphi(\bss)$ and $\omega(\bss)$
should go to zero at infinity as ${(\bss)^{-1/2}}$. If we introduce 
this ansatz into the Bogomol'nyi equations, we find that
\beq
\omega(\bss) = \tau\varphi(\bss) ~,
\label{propo}
\eeq
with $\tau$ an arbitrary real constant, and
\beq
r\frac{d\varphi}{dr} + \xi\varphi = 0 ~,
\label{ans1}
\eeq
\beq
\frac{1}{r}\frac{d\xi}{dr} + \frac{2e^2v^2}{\kappa^2}(r^2 + 
\tau^2)\varphi^2[1 - (r^2 + \tau^2)\varphi^2] = 0 ~.
\label{ans2}
\eeq
The function $\xi$ must vanish at the origin in order to have a 
well-defined solution. Indeed, it is easy to see that it vanishes
as $\xi \approx r^2$, thus leading to a non-zero magnetic field at
the center of the vortex. The same result was previously obtained 
in the non-supersymmetric case \cite{K}, and it is worth to stress
on the fact
that this behaviour is different from that of the abelian selfdual
Chern-Simons vortex \cite{HKP,JW}. On the other hand, it is easy to 
compute the asymptotic behaviour of the magnetic field, ${\cal F}
\approx r^{-4}$, which is also different from the exponential decay
of the abelian selfdual solution. Our solution is labeled by the
real parameter $\tau$. There is also an arbitrary relative phase
in the definition of the basis ($\Phi_1$,$\Phi_2$).

~

To end this letter, we would like to comment on the fact that the 
transformations generated by the antichiral parameter $\eta_-$
are broken in the background of the BPS vortex
\beq
\delta_{\eta_-}\Psi\vert_{\mbox{\footnotesize BPS-vortex}} = 
\frac{2e^2}{\k}(\pdp - v^2)\Phi\,\eta_- + D_{\ze}\Phi\,\eta_+ \neq 0 ~.
\label{broken}
\eeq
These variations give zero energy Grassmann variations of the vortex 
configuration; that is, they are zero modes of the Dirac equation in 
the background of the supersymmetric Chern-Simons semilocal vortex. 
Quantization of these fermionic Nambu-Goldstone zero modes leads
to the construction of a BPS supermultiplet of degenerate bosonic
and fermionic soliton states which transform according to a short
representation of the supersymmetry algebra.

~

In conclusion, we have considered in this letter the $N=2$ supersymmetric
SU$(2)_{\mbox{\tiny global}}\times$U$(1)_{\mbox{\tiny local}}$ 
Chern-Simons-Higgs
system. We have shown that the requirement of $N=2$ supersymmetry forces 
a special relation between coupling constants and dictates a specific 
shape for the Higgs potential while, at the same time, through the 
supercharge algebra, it imposes the Bogomol'nyi equations on certain 
classical field configurations, identified as the BPS semilocal 
Chern-Simons cosmic strings. They are stable --as a consequence 
of the underlying $N=2$ supersymmetric invariance-- even though 
the manifold of the Higgs vacuum for this system does not contain 
non-contractible loops. 

~

A more suitable model, from the point of view of cosmology, would 
require the coupling to gravity. In that context, semilocal cosmic string
and multi-string solutions were explicitly found and studied a few
years ago \cite{GORRS} for the model of Ref.\cite{VA}. It would
be of great interest to seek such kinds of solutions in the
semilocal CSH model coupled to gravity. In particular,
it would be interesting to know if the coupling of these BPS semilocal
Chern-Simons vortices to gravity leads to the existence of Killing 
spinors in asymptotically conical spacetimes, in the spirit of the
solutions studied earlier in Ref.\cite{JDE}. Even the coupling of local 
BPS Chern-Simons vortices to N=2 supergravity seems to be a problem of
great interest. It is well-known \cite{PV,CL,GL,GC} that self-dual gravitating 
solitons in purely bosonic Chern-Simons systems require an eight-order
Higgs potential. Whether such a behaviour is forced or not by extended 
supersymmetry is not known. In particular, if the answer is affirmative, 
it is quite puzzling that supersymmetry forces qualitatively different Higgs 
potentials provided it is a local or a global symmetry. In this respect,
it has been recently suggested \cite{GFMG} that the flat metric limit of an 
N=2 supergravity theory whose bosonic sector is the CSH model coupled to 
gravity \cite{PV,CL,GL,GC}, could be given by a generalized N=2 supersymmetric 
CSH theory that, as shown in Ref.\cite{GFMG}, admits 
self-dual solitons for an eight-order 
Higgs potential. Clearly, there is much scope for further study.
\vskip 3mm

This work has been partially supported by a Spanish Ministry of 
Education and Culture fellowship.

\end{document}